# Diagnostic systems in DEMO: engineering design issues

T N Todd

*EURATOM/CCFE Fusion Association, Culham Centre for Fusion Energy,
Culham Science Centre, Oxfordshire, OX14 3DB, UK*

**Abstract.** The diagnostic systems of DEMO that are mounted on or near the torus, whether intended for the monitoring and control functions of the engineering aspects or the physics behaviour of the machine, will have to be designed to suit the hostile nuclear environment. This will be necessary not just for their survival and correct functioning but also to satisfy the pertinent regulatory bodies, especially where any of them relate to machine protection or the prevention or mitigation of accidents foreseen in the safety case. This paper aims to indicate the more important of the reactor design considerations that are likely to apply to diagnostics for DEMO, drawn from experience on JET, the provisions in hand for ITER and modelling results for the wall erosion and neutron damage effects in DEMO.

## INTRODUCTION

The world fusion community is increasingly focusing on the practicalities entailed in constructing and operating the first demonstration fusion reactor, DEMO, stimulated by the ongoing considerations of building ITER and of course its suite of plasma diagnostics and control systems. Many scientists new to the largest (and now the only DT capable) operating tokamak JET find the engineering requirements imposed upon all assemblies comprising or fixed to the tokamak load assembly – including all the plasma diagnostics - to be surprisingly onerous compared to those they have experienced on other fusion research machines, which apart from TFTR will have been non-nuclear. However JET is not subject to licensing by any nuclear regulatory body, although of course it complies with the requirements of the UK Health and Safety Executive and the UK Environment Agency.

JET is presently the nearest machine to DEMO that is in operation, while ITER is of course a vital intermediate step. JET operates under a rigorously applied safety case following the same principles as those developed for the fission plant that the JET Operator (the UK Atomic Energy Authority) once operated. Accordingly diagnostic systems mounted on the machine have to satisfy over twenty criteria, described in a JET design approval check-list [1]. These design constraints and approvals, necessary to achieve the implementation of any diagnostic system on JET, are necessarily more onerous than those applied to other (non-DT) operating tokamaks. It should be noted that DT operation in JET involved a site inventory of only 20g of tritium, totalled about $3 \times 10^{20}$ neutrons and resulted in torus radiation fields two weeks after shut-down of around 10mGy/hr, while for ITER the figures are 4kg, $\sim 3 \times 10^{27}$ neutrons and 500Gy/hr, and DEMO perhaps 6kg, $\sim 1.4 \times 10^{29}$ neutrons (in $\sim$3FPY) and about 20kGy/hr. The associated neutron damage (which loosely corresponds to the neutron fluence but is dependent upon the material involved and the neutron spectrum), for components near the plasma, is in the region of $1.2 \times 10^{-6}$, 2 and 60 displacements per atom respectively. It can immediately be seen that as onerous as the JET design criteria might be, to suit its nuclear context, those of ITER and one day DEMO are going to be far more so.

In addition to the nuclear considerations of components intended to be mounted in or on the DEMO vacuum vessel, there are also likely to be problems of erosion of components with a line of sight to the plasma, due to energetic charge-exchanged neutrals, competing with deposition of condensable material eroded from elsewhere in the machine or synthesised in the low temperature edge regions of the plasma and its surrounding gas and vapour. Here the differences from present day machines are largely due to the near-continuous operation of DEMO, compared to the very low duty cycle now typical.

The full set of engineering requirements for diagnostic systems implemented in DEMO sets the stage for another important discussion, yet to be resolved, regarding the degree of sophistication that the DEMO diagnostics should achieve. Will DEMO have sufficient new physics (e.g. due to operation in some special mode requiring precise profile control yet with the alpha power dominating over the auxiliary heating and current drive power, and with an unusually high radiated power fraction) to demand a very comprehensive and high resolution diagnostic set? Or must it operate in a robustly simple, perhaps less efficient mode that is consistent with very limited, low resolution plasma characterisation? Or is there a compromise, in which sophisticated diagnostics are installed without the costly features ensuring long life and maintainability in the hostile DT environment, and only permit detailed checking of physics for He, H and DD operation and the early phases of DT commissioning, thereafter trusting to simpler diagnostics and integrated modeling for machine control?

## ENGINEERING REQUIREMENTS

Presently operating tokamaks (and other magnetic confinement devices) vary in the engineering requirements that must be met by the design of the associated plasma diagnostic systems. Most of them require UHV compatibility and resistance to mechanical damage from the transient accelerations caused by plasma disruptions, often with arrangements to minimise optical component degradation by accumulated deposits of material recycled by the plasma. Some acknowledge the problems of eddy and halo currents induced during disruptions and vertical displacement events, and of course JET requires features to guarantee high integrity of tritium (and beryllium dust) containment. However any fusion reactor will have far more onerous engineering requirements for all systems connected to the torus, ultimately to satisfy the nuclear and environmental regulatory bodies of the host country regarding the human safety and environmental protection of the installation. Table a) lists some of the more obvious design considerations that will be necessary, the non-nuclear ones being commonly addressed in present tokamaks.

**TABLE a).** Examples of probable considerations in the design of DEMO plasma diagnostics

| Type | Compatibility/requirement |
|---|---|
| Non-nuclear | avoidance of halogens (which can form acids in tritium plant) |
| Non-nuclear | earthing and signal paths |
| Non-nuclear | EMI screening (source or recipient) |
| Non-nuclear | disruption and seismic acceleration forces |
| Non-nuclear | vibration and other operational cyclic loads |
| Non-nuclear | disruption-induced currents and voltages |
| Non-nuclear | EMI screening (source or recipient) |
| Non-nuclear | ICRH and ECRH immunity |
| Non-nuclear | vignetting of other systems (photon, particle) |
| Non-nuclear | UHV design principles |
| Non-nuclear | fast particle impacts (i.e. erosion, heating) |
| Non-nuclear | deposition of dust and coatings on critical surfaces |
| Nuclear | real-time radiation-induced effects |
| Nuclear | radiation damage (e.g. in mirrors and transparent optics) |
| Nuclear | compatibility with rapid remote handling systems |
| Nuclear | useful life between component replacements (if possible) |
| Nuclear | neutron streaming (labyrinths, line of sight restrictions) |
| Nuclear | tritium and active dust containment |
| Nuclear | use of reduced activation materials and coolants |
| Nuclear | installation and operational clearances |
| Nuclear | impact on neutron fluxes elsewhere |
| Nuclear | thermal environment including self nuclear heating |
| Nuclear | RAMI (reliability, availability, inspectability, maintainability) |

ITER represents a very important interim step between JET and DEMO, and inspection of the ITER "Dynamic Object-Oriented Requirements System" (DOORS) [2] documentation for the plasma diagnostics reveals a number of considerations new to fusion research machines and somewhat demanding – but nevertheless very much easier to accommodate in many respects than those for DEMO will be, due to the implications of its larger size and considerably increased nuclear performance.

One example from the ITER Diagnostics DOORS is the need to determine the normal and off-normal operational environment of each sub-assembly within the machine, e.g. local temperature, peak magnetic field and rate of change of magnetic field, and acceleration loads (seismic as well as disruption-induced). In JET the engineering analyses to demonstrate adequate lifetime of any such sub-assembly begin simply by identifying the zone in the tokamak load assembly in which the sub-assembly lies. The peak magnetic fields and rates of change of field (separated into poloidal and radial components) and acceleration vectors (for example) are defined with respect to poloidal and radial location, providing worst cases to form the basis of the analyses, a very much streamlined approach compared to assessing the loads imposed on each sub-assembly uniquely.

ITER will be the first fusion research machine to impose significant nuclear heating on components seeing any significant part of the first wall neutron flux. Thus cooling provisions need to be made for any such components, even if these are only to ensure sufficient good thermal contact with other structures with active cooling – some of which, such as the vacuum vessel, will only be cooled to ~150°C or more. DEMO will have very much higher operational temperatures, in order to achieve a high Carnot efficiency from the coolant. Even when DEMO has been shut down for some weeks, active cooling will be necessary to inhibit the neutron-induced decay heat from raising the temperature of components (especially any with a small surface area to volume ratio) to several hundred °C, complicating the (inevitably remote handling) maintenance activities.

## NEUTRON IRRADIATION EFFECTS

### Continuous or Real-Time Effects

Neutron wall loading is often stated as a simple average, i.e. the total fusion neutron power divided by the total surface area of the first wall, which is easy to estimate for any given size and power of a fusion reactor and readily re-interpreted in terms of the flux of neutrons/$m^2$/sec. However when considering the effects produced by the neutron interactions with matter, the flux in all directions and in all neutron energies has to be considered. Even the raw 14MeV neutrons interact via their track length in the surface of the first wall, the associated geometric effects raising the flux by a factor of ~1.6 for typical fusion tori. More importantly, the wall and blanket structures scatter the neutrons considerably as part of the thermalisation process, while the presence of a neutron multiplier (Pb or Be) increases the neutron flux by ~30%. The result of all this is to raise the effective all-directions flux of damaging neutrons (>0.1MeV) at the first wall by a factor ~10 compared to that of the raw 14MeV neutrons. This should be borne in mind when considering placing diagnostic components in the first wall environment.

Energetic neutrons are an example of ionizing radiation, and some of the real-time effects are due to excitation of electrons, e.g. taking them out of the closed inner shells of insulators and into the conduction band, creating Radiation Induced Conductivity [3,4] as summarised in figure 3 of [3]. Items at or near the first wall in DEMO in operation might be expected to see ~$10^3$-$10^4$ Gy/s, reducing the resistivity of inorganic insulators to ~10MΩm, marginal for high impedance circuits or low tan-δ (RF absorption) applications. In addition, there can be Radiation Induced EMF and Radiation Induced Thermal EMF [5,6], the latter due to differential heating of any dissimilar metal joints in detector circuit paths. These are likely to be significant for low-voltage signals such as those from thermocouples or magnetic pick-up coils in slowly varying fields.

Another class of real-time interference in diagnostic signals is that of the direct neutron interaction, giving rise to such phenomena as scintillation in optical materials (such as lenses and fibre optics), and charge redistribution in detectors (CCD pixels, photomultipliers, channeltrons etc). Such effects could easily lead to "white-out" of any CCD cameras subject to significant radiation fluxes, even if such cameras were made to survive the accumulated neutron damage by some arrangement providing only fleeting exposure.

### Accumulated Damage Effects

Most of the real-time effects described above are familiar problems in present day DD machines of high performance, while white-out or "snow-storm" degradation of video camera signals was a feature of DT operation in JET. However in ITER to some extent, and in DEMO to a large extent, the total neutron fluence in the diagnostic life between (any) replacements will be of a magnitude completely new to fusion experience. The accumulated damage in conductors and insulators will steadily degrade the diagnostic performance and may result in complete failure of individual channels.

Well-known from TFTR and JET is the problem of darkening of transmissive optical components due to the development of colour centres in the material lattice, caused by displacement of lattice ions or transmutation of the atomic elements. At moderate levels, this darkening can be annealed out by taking (or holding) the material at an elevated temperature, typically >300°C, but annealing cannot correct elemental transmutation, so residual darkening will slowly accrue [6].

Lattice ions suffering a near head-on impact by fast neutrons will recoil with significant energy (e.g. up to several hundred keV for iron), then ploughing through the lattice for many hundreds of interatomic lattice spacings, creating cascades of secondary recoil ions so that thousands of lattice ions are displaced into interstitial lattice sites. Nearly all of the resulting "Frenkel Pairs" (i.e. interstitial atoms and lattice vacancies) recombine within picoseconds (and faster and more completely at higher temperature) but dozens to hundreds remain for each recoil ion traversal, constituting lattice defects. Such defects generally raise the yield strength of metallic materials but embrittle them, risking mechanical failure if a degree of ductility (plastic compliance) was a necessary feature of the diagnostic assembly in operation. Alloys differ in their sensitivity to neutron fluence in this regard, as shown for copper in slide #4 of [7] and steel in figure 4 of [8]. These references both show that the effect is well characterised by the number of displacements per atom suffered by the metal, and not the hardness of the neutron spectrum, since fission, fusion and spallation data are included and have the same trends.

In fission systems, the virgin neutron spectrum resembles a pseudo-Maxwellian of ~2.3MeV "temperature" and the core assembly moderates this strongly (especially in thermal fission reactors, of course). As a result, relatively few of the neutrons have energies above the transmutation thresholds of the materials used in the structure (and diagnostics) of the reactor, since these are mostly above ~8MeV. In DT fusion systems however, there is a substantial population of neutrons above this typical threshold energy, resulting in a much greater ratio of transmutation to displacement damage (~1 atomic part per million or appm He per dpa for a fission reactor, ~8appm He per dpa for DEMO) [9]. Most of the gaseous transmutation products are protons but ~10% are alpha particles, He. The hydrogen diffuses out of the metal fairly readily but the helium does not, developing bubbles in radiation induced voids that tend to accumulate at the lattice grain boundaries. These strain the lattice and cause macroscopic swelling of the material, summarised in slide 8 of [10] and figure 6 of [11], with obvious deleterious effects on the overall assembly, especially if the gradient of neutron fluence in the assembly is significant. In addition, the presence of the helium makes it difficult to achieve rewelded joints of acceptable quality, with ~1 appm of helium representing a typical practical upper limit to the helium concentration [12].

Another adverse effect of neutron fluence is weakening of insulators, especially organic types. This is generally taken to be a dose effect and is therefore accounted for in Grays, with doses ~50-200MGy bracketing the upper limit of typical organic insulators such as epoxy resin (low) or cyanate ester (high), at which interlaminar shear strength has been degraded by ~30%. Mixtures of these two types offer intermediate radiation tolerance and ease of fabrication techniques such as vacuum impregnation [13, 14].

Although the neutron spectrum depends upon the reactor design and materials vary in the relative magnitudes of these damage effects, very roughly it can be taken that 1MW-yr/m$^2$ of average neutron wall load corresponds to a total flux of ~1.4x10$^{26}$n/m$^2$, ~8dpa, ~80 atomic parts per million of He and ~(10$^4$-10$^5$) MGy. If organic insulators are present, these will usually set the operational limit. Copper suffers significant radiation-induced loss of conductivity, around 50% at 60dpa [11] and not entirely removed by annealing, and considerable swelling as noted above, and so is likely to be the next most easily damaged part of an electronic circuit or heat sink assembly.

## EROSION/DEPOSITION EFFECTS

Many plasma diagnostic systems in operation on current fusion devices use mirrors, lenses or windows close to the edge of the plasma, often with shutters closed during glow discharge cleaning to avoid the deposition of coatings that degrade the transmission. In a high performance, steady state (or pulsed but high duty cycle), fusion reactor, the erosion rate of the first wall (and therefore any object at an equivalent location) by sputtering is difficult to forecast accurately but is likely to represent a considerable challenge. It is difficult to forecast because it is a very strong function of the mass and characteristic energy of the incident species, as discussed at length and epitomised by figures 7 and 8 of [15] for tungsten as the target material, which suggests erosion rates varying across several orders of magnitude from a fraction of a mm to many cm per full-power year of reactor operation. Tungsten is selected for the first wall because it has a high elemental mass (making it resistant to physical sputtering) and good thermomechanical behaviour: lenses and windows will generally not meet those criteria, although mirrors can. Even for mirrors, any grain boundaries in the optical surface represent weaknesses subject to accelerated erosion, so single crystal mirrors are necessary to preserve a good optical performance.

The incident fluxes of fuel, ash and impurity species introduced to enhance radiative cooling of the plasma (i.e. to minimise power loading on the divertor) depend upon the edge physics behaviour of the machine and cannot reliably be predicted with the present level of understanding of the plasma periphery. Estimates for the neutral particle flux emerging from the main plasma are expected to lie in the $10^{17}$-$10^{19}$ atoms/cm$^2$/sec range, while the temperature will relate to that of the edge pedestal region of the main plasma, unless a low temperature edge region of high line density could be provided outside the pedestal to dictate the typical energy of the escaping charge-exchange neutrals. The present level of understanding of edge physics phenomena does not allow a robust prediction to be made that such a cold mantle adjacent to a high temperature pedestal could be stably provided, however.

Clearly, unlike in the charged particle flow to the divertor plates, the impurity species must be entirely neutral to reach the first wall and in general have a charge state in the plasma edge which is above unity, hence requiring more than one charge exchange event (primarily on a different species, i.e. via a non-resonant process) to become neutral. This implies that the first wall neutral particle flux will be dominated by the easily charge-exchanged hydrogenic fuel species. If ELMs occur, however, and can perturb the plasma boundary sufficiently for the associated plasma filaments to reach the wall, then as implied by the analyses of [15], the associated fluxes of impurity ions (at temperatures characteristic of the pedestal) are likely to add to the erosion rate significantly.

The annual erosion of the first wall could therefore lie anywhere in the range of tens of microns to perhaps tens of millimetres, with the latter a significant challenge for maintaining the integrity of the first wall, let alone any precision diagnostic components in its vicinity. Accordingly it would be prudent to locate all plasma diagnostic components well away from the first wall, preferably behind the blanket and shield, and preferably seeing the plasma (if necessary) via narrow labyrinths with no mirrors near the plasma.

A possibly more tractable problem for mirrors and windows etc. is that of deposition (the mitigation of which is discussed below). This could be dust or condensate of material sputtered or evaporated from the first wall and divertor assemblies, or layers created by glow cleaning plasmas or plasma chemistry associated with the plasma edge and is interactions with surrounding surfaces. In present tokamaks it is dominated by hard dark layers composed of carbon rich compounds and/or metallic deposition from gettering or plasma recycling of wall and in-vessel component materials, but the much larger erosion between maintenance periods due to the near-continuous plasma operation of DEMO, and possibly higher particle fluxes and energies, may make some form of dust an important contribution to optical component degradation. As with erosion, a natural preventative solution is to ensure that all critical diagnostic surfaces are located well away from the plasma, protected from all plasma interactions and if possible also from dust accumulating in the main torus.

## IMPACT ON TRITIUM BREEDING RATIO

Although a continuous volume of lithium-bearing breeding material completely surrounding the plasma and several metres thick could achieve a tritium breeding ratio (TBR) of around two, the design of a breeding blanket in a practical tokamak reactor necessarily incorporates many features that reduce this figure considerably. The successive effects of introducing a sequence of very plausible practicalities were thoroughly assessed in [16], figure 5 therein showing that even with a neutron multiplier and considerable enrichment of the natural lithium with $^6$Li (to make better use of the thermalised neutrons), it is quite difficult to achieve a TBR of 1.05, while 1.15 is often a design target to allow for uncertainties in the nuclear data and the eventual engineering implementation. It should be noted that in this study, the first wall area given over to diagnostics was confined to the outboard wall and only ~0.7% of the area of that, to be compared to nearly 20% of the outboard vessel area in ITER, while the thickness of tungsten cladding in the first wall was taken to be 10mm in the important outboard mid-plane region and 50mm in narrow strips (to enhance vertical plasma position control) well away from the mid-plane. Other studies have shown that changing the thickness of the tungsten cladding on the first wall from 2mm to 20mm (probably necessary to accommodate a high rate of erosion) would reduce the TBR by about 15% [17], putting further pressure on the other design aspects, including reducing the breeding area lost to diagnostics. Unsurprisingly, if divertor power loading or erosion rates led to the adoption of a double-null divertor configuration, the further loss of effective tritium breeding area would pull the TBR down by ~20% [17]. As a result of these TBR constraints, it is considered unfeasible to put ITER-style diagnostic port plugs into DEMO, and the DEMO plasma diagnostics will have to be compatible with very narrow sight lines or be of types that do not require a line of sight to the plasma and can operate effectively behind the blanket modules or even outside the vacuum vessel. However consideration of the design freedom available to the blanket designers, as discussed in [16], shows that increasing the blanket thickness or $^6$Li enrichment (or any of several other key parameters such as neutron breeding material, coolant type, percentage of blanket given over to coolant) would permit a solution to be found if it was considered necessary to allocate a relatively large port area to diagnostic systems – it becomes an issue of whole-life net cost analysis, which should in any case always be paramount in machine design considerations.

## REMOTE HANDLING ISSUES

The problem of nuclear heating during full power burn in DEMO was alluded to above, as was the issue of nuclear decay heating during shut-down periods, especially when the blanket modules are nearing end of life. The latter could take the temperature of large components such as first wall modules to around 1000°C once the coolant flow has been interrupted, reducible to perhaps ~800°C if temporary gas cooling is provided to facilitate the remote handling of such components [18]. Similar temperatures could be expected to be suffered by plasma diagnostic systems embedded in the blanket modules but the present intention is to commit these to waste stores and disposal (with appropriate segregation where possible), without removing any sub-assemblies for re-use. The key point is that the decay heating can be significant in any activated assembly and if active cooling of some diagnostic system was necessary in plasma operation, it will probably need to be continued in shut-downs and possibly during any maintenance activities, inevitably involving remote handling.

In ITER, many of the diagnostic systems can only be replaced at the same time as, or even simply as part of, the first wall shield modules or divertor cassettes. A similar policy is very likely to pertain for DEMO diagnostics, so that the minimum design life would be set by the replacement time of those assemblies, typically a few years. Certain types of remote handling task might be conceived to allow plasma diagnostics to continue to function reliably over the intended life between replacements, perhaps for cleaning, replacement or renewal of critical surfaces, *in situ* calibrations or exchange of exposure tokens etc. At present, neither in ITER nor DEMO is there any intention to provide any small, very rapidly deployable in-torus or in-port manipulator to effect such maintenance activities – there are only the large scale remote handling devices for replacing wall modules and divertor cassettes (and possibly in ITER the MultiPurpose Deployer, but even that is not very small or capable of deployment in a timescale less than several weeks). If any maintenance activity is needed for a diagnostic between its major replacements, then, it will be necessary for the designer to provide a suitable remote handling system to do it, and above all this must be both fast and reliable, with a robust failure mode recovery strategy, both for itself and whatever it was supposed to be maintaining. The alternative is loss of availability (operating time) for the reactor, costing ~£€1M per day in lost electricity sales at today's pool price for a 1GW-electrical power plant.

## POSSIBLE SOLUTIONS

Clearly the neutron flux and fluence and their effects will fall off with distance through the blanket and shield of the reactor, indeed the damage (dpa) by a decade in about 0.4m of blanket thickness (depending on the blanket materials) and the helium production rate by about two decades per 0.4m [19] (since the neutron spectrum softens with distance through the blanket). Thus if diagnostics can serve a useful function when placed behind the blanket, their lifetime will be greatly extended compared to that of similar systems located at or near the front of the blanket modules. Naturally diagnostics based upon penetrating radiation (neutrons and gamma rays) can be located outside the vessel and still provide useful information on the plasma behaviour. The neutrons will readily stream down port tubulations or any gaps, such as those between blanket modules, so intermediate fluxes and spectral characteristics are to be expected in and near such channels. A study undertaken to assess the remote handling feasibility, including reweldability of steel pipes, of structures in and near the divertor and its access duct [20], suggests that in the duct, the reweldability criterion (of 1appm of He content in the steel) is only likely to be satisfied at about 5m away from the edge of the plasma, where during operation the neutron flux will be only $~1.5 \times 10^{17}$n/m$^2$-sec.

Although conventional lenses (or windows) near the plasma would be likely to have a rather short life, even if operated at several hundred °C to inhibit internal darkening and noise suppression signal processing to reduce scintillation effects, the option of Zone Plate or "Photon Sieve" lenses (see, e.g. figure 16 on page 156 of [21]), diffractive by nature, should be investigated for this type of harsh environment. Similarly although conventional diffraction gratings would be likely to degrade rapidly when exposed to erosion or deposition effects, a class where the reflective strips are formed on long supporting web plates may offer a degree of immunity to surface coatings or erosion. An example is given in the eighth figure of [22]. Depending on the application, the type of structure shown here might benefit from blazing (i.e. tilting the reflective surface to concentrate the diffracted light into the desired order of diffraction), polishing of the reflective surface and fabrication in metal rather than a metal-coated dielectric.

Where mirrors and lenses seem to be inescapable, it may be possible to clean them in *in situ* using laser ablation techniques, such as those demonstrated in [23, 24]. This may provide a back-up solution to deal with the residual deposition effects if such optical components are shuttered and only exposed to the plasma etc. with a very low duty cycle, as mentioned below.

A fast transfer system (also known as a "rabbit" system since it comprises rabbit-sized containers moving rapidly in small labyrinthine tunnels) might be considered to enable certain types of compact diagnostics or optical assemblies to be used in otherwise impossibly hostile situations, or to confer a degree of time resolution to detectors usually only considered as inter-maintenance flux integrators, such as passive irradiation tokens. Any such system would need to be convincingly reliable, however, especially if the resulting measurements were key to the operation or safety of the plant. Along similar lines, but again demanding high reliability of complex mechanical systems in a very hostile environment, vulnerable diagnostics could in principle be located behind shutters and/or movable shield blocks to minimize their exposure to the various deleterious effects outline above.

Another option, beginning to be explored in ITER, is to acknowledge that some diagnostics should be installed in locations permanently inaccessible to maintenance devices, or destined to fail in service before the first planned shutdown for a maintenance intervention, but to use their more sophisticated or more comprehensive measurements to validate one or more codes demonstrating robust control of the plasma based solely on the signals obtained from a more limited set of reliable, long-life and readily maintainable diagnostic systems. Then the anticipated failure of the more sophisticated systems could be demonstrated not to compromise the safe and cost-effective operation of the machine, while some valuable information adding to understanding of the machine behaviour would be obtained during the non-radioactive and initial low fusion power commissioning operations.

## CONCLUSIONS

This study has endeavoured to recognise those aspects of fusion reactor design that are challenging for the principal components of the reactor in terms of the operational demands on temperature, surface erosion and deposition effects, and the continuous and accumulated neutron damage, in order to consider the implications for the plasma diagnostics necessary to control the machine properly and thus guarantee the safety of its operators and the wider public. Regulatory concern will focus on the demonstration of adequately comprehensive and reliable (hence accessible for inspection, calibration and maintenance) diagnostics, actuators and associated control codes, while commercial considerations will add a need for very high availability and hence very much faster techniques of remote handling for all maintenance activities than used on JET or foreseen for ITER. Even when shut down, after significant full power operation, the decay heat and gamma dose rates are likely to create significant challenges for the intervention activities, being many orders of magnitude higher than those of TFTR, JET and ITER. Accordingly electronic devices normally used for vision systems and force feed-back in remote handling may suffer limited lifetimes and/or considerable interference.

Components intended to remain in the reactor for the many years foreseen between the major interventions for divertor and first wall replacements will have to survive the neutron fluence pertinent to their location, which creates a range of deleterious effects capable of causing change of sensitivity, alignment and perhaps failure. The real-time neutron and gamma fluxes introduce many effects contributing to signal contamination which need to be minimised by design and/or reliably compensated in order to secure appropriately accurate control of the plasma.

In this light, it becomes clear that if it is essential to operate DEMO with finely structured plasma profiles in order to achieve adequate stability or fusion gain, it will be especially challenging to meet all the necessary design and implementation criteria in providing the associated high resolution and sophisticated plasma diagnostics and their inspection, calibration and maintenance systems. An interesting option, if acceptable to the nuclear regulators, is that of providing a sophisticated diagnostic set expected to survive in full only for the early commissioning phases of the machine, making use of the resulting data to validate control codes based on a much simpler set of diagnostics designed to be able to survive the hostile environment. Ultimately, though, the degree of sophistication desired in the measurement and control of the ongoing operation of the reactor should be a question of overall whole-life cost, or perhaps the amortised cost of each unit of electricity sold to the utilities. Whatever the diagnostics philosophy though, the preparations for the implementation of plasma diagnostics in DEMO, right from the present early concept stage, must recognise the degree of rigour and formal regulatory scrutiny represented by all the engineering design requirements for the full life of the machine.

## ACKNOWLEDGMENTS

I gratefully acknowledge the help of V. Riccardo, R. Lobel, S. Gerasimov, N. Hawkes, S. Popovichev, A. Widdowson and K-D. Zastrow for JET-related information and references, M.J. Walsh and G. Vayakis for pointing me to the ITER *DOORS* document for plasma diagnostics, and A. Loving for aspects of diagnostics remote handling in DEMO. I also gratefully acknowledge the insights into the problems of implementing plasma diagnostics in DEMO that can be found in the internet, particularly the presentations made available by K. Young and T Donné.

This work, part-funded by the European Communities under the contract of Association between EURATOM/CCFE was carried out within the framework of the European Fusion Development Agreement. For further information on the contents of this paper please contact PublicationsManager@ccfe.ac.uk. The views and opinions expressed herein do not necessarily reflect those of the European Commission. This work was also part-funded by the RCUK Energy Programme [grant number EP/I501045].